\documentclass[journal,letterpaper]{article}

\usepackage[ansinew]{inputenc}
\usepackage[T1]{fontenc}

\usepackage{amsmath,amssymb,amsfonts,latexsym,amsthm,bm}
\usepackage{mathrsfs}
\usepackage{dsfont}
\usepackage{upgreek}
\usepackage{cancel}
\usepackage{breqn}

\usepackage[none]{hyphenat}
\usepackage{fancyhdr}
\usepackage[top=1.5cm,bottom=1.5cm,left=1.5cm,right=1.5cm]{geometry}

\usepackage{algpseudocode}
\usepackage{algorithm}

\usepackage{graphicx}
\usepackage{graphics}
\usepackage[normal]{subfigure}
\usepackage{float}

\usepackage{booktabs}
\usepackage{colortbl}
\usepackage{ctable}
\usepackage{multirow}

\usepackage{courier}
\usepackage{listings}

\usepackage{color}
\usepackage{mdframed}
\mdfsetup{backgroundcolor=brown!10,roundcorner=10pt}
\mdfdefinestyle{mdfexample1}{leftmargin=1.5cm,rightmargin=1.5cm,
	innertopmargin=7pt,innerleftmargin=1cm,innerrightmargin=1cm}
\mdfdefinestyle{mdfexample2}{leftmargin=0.5cm,rightmargin=0.5cm,
	innertopmargin=0pt,innerleftmargin=1cm,innerrightmargin=1cm}
\definecolor{myblue}{rgb}{0.64, 0.88, 1}
\definecolor{myyellow}{rgb}{1, 1, 0.8}
\definecolor{mygray}{rgb}{0.9, 0.9, 0.95}

\usepackage[toc,page]{appendix}

\usepackage{balance}
\usepackage{enumerate}

\usepackage{cite}
\usepackage{url}

\usepackage{soul}

\theoremstyle{plain}



\begin{document}


\title{ \vspace{-1cm} {\LARGE { \textbf{Sparse Linear Models applied to Power Quality Disturbance Classification}}
} \vspace{-0.6cm} }
\author{ \normalsize Andrés F. López-Lopera, Mauricio A. Álvarez, and \'Avaro A. Orozco.\\
		 \normalsize Faculty of Engineering, Universidad Tecnol\'ogica de Pereira, Colombia, 660003.
		}
\markboth{Universidad Tecnológica de Pereira. Maestría en Ingeniería Eléctrica. \today}%
{Shell \MakeLowercase{\textit{et al.}}: Bare Demo of IEEEtran.cls for Journals}
\date{ \vspace{-0.6cm} \normalsize \today}
\maketitle

\sloppy
\vspace{-1.2cm}

\newcommand{\THD}{\mbox{THD}}
\newcommand{\Bx}{\textbf{x}}
\newcommand{\Ba}{\textbf{a}}
\newcommand{\BF}{\textbf{F}}
\newcommand{\BS}{\textbf{S}}
\newcommand{\BX}{\textbf{X}}
\newcommand{\By}{\textbf{y}}
\newcommand{\BY}{\textbf{Y}}
\newcommand{\Bz}{\textbf{z}}
\newcommand{\BZ}{\textbf{Z}}
\newcommand{\Bw}{\textbf{w}}
\newcommand{\BW}{\textbf{W}}
\newcommand{\Bbeta}{\boldsymbol{\beta}}
\newcommand{\Bell}{\boldsymbol{\ell}}
\newcommand{\Balpha}{\boldsymbol{\alpha}}
\newcommand{\Bmu}{\boldsymbol{\mu}}
\newcommand{\BPhi}{\boldsymbol{\Phi}}
\newcommand{\BPsi}{\boldsymbol{\Psi}}
\newcommand{\BSigma}{\boldsymbol{\Sigma}}

\vspace{15pt}
\begin{abstract}
	Power quality (PQ) analysis describes the non-pure electric signals that are usually
	present in electric power systems. The automatic recognition of
	PQ disturbances can be seen as a pattern recognition problem, in
	which different types of waveform distortion are differentiated
	based on their features. Similar to other quasi-stationary
	signals, PQ disturbances can be decomposed into time-frequency
	dependent components by using time-frequency or time-scale
	transforms, also known as dictionaries. These dictionaries are
	used in the feature extraction step in pattern recognition
	systems. Short-time Fourier, Wavelets and Stockwell
	transforms are some of the most common dictionaries used
	in the PQ community, aiming to achieve a better signal
	representation. To the best of our knowledge, previous works
	about PQ disturbance classification have been restricted to
	the use of one among several available dictionaries. Taking
	advantage of the theory behind sparse linear models (SLM), we
	introduce a sparse method for PQ representation,
	starting from overcomplete dictionaries. In particular, we apply
	Group Lasso. We employ different types of time-frequency (or
	time-scale) dictionaries to characterize the PQ disturbances, and
	evaluate their performance under different pattern recognition
	algorithms. We show that the SLM reduce the PQ classification
	complexity promoting sparse basis selection, and improving the
	classification accuracy.
\end{abstract}

\section{Introduction}
Power quality (PQ) and reliability, and equipment diagnostics, control and protection are some of the case studies of signal processing in electric power systems \cite{Bollen2009}. PQ studies evaluate the waveform of voltage and current signals respect to pure sinusoidal waveforms with a single frequency component (e.g. 50 or 60 Hz) \cite{Surajit2011, kusko2007}.  Non-linear electronic devices such as cycloconverters, alternators, and current starters \cite{kusko2007, Pasko2012} introduce undesirable harmonic distortions to the network. Other examples of sources that distort pure sinusoidal electric signals include electrical loads with high values, electrical faults on the network \cite{kusko2007}, tap changer \cite{Farazmand2014}, and capacitors switching banks \cite{Pasko2012}. Depending on the source of PQ disturbances, we can find distortions such as swell/sag, harmonics, flickers, notching and transients \cite{Bollen2009, kusko2007, Surajit2011}. 

The automatic recognition of PQ disturbances is an important research field in PQ analysis. It can be stated as a pattern recognition problem, in which the different types of waveform distortions are differentiated based on their features \cite{Granados}. For the feature extraction, signal processing practitioners in PQ, have employed several basis function superposition, also known as dictionaries \cite{Chen1998}. Similar to other quasi-stationary signals, PQ disturbances can be decomposed into time-frequency dependent components by using time-frequency or time-scale dictionaries \cite{Bollen2009, Granados}. Short time Fourier transform (STFT) \cite{Sharma2013}, Wavelets transform (WT) \cite{Vega2007, Eristi2012}, and S-transform (ST) \cite{Naik2011, Huang2009} are commonly used in PQ representation for the feature extraction. After the feature extraction step, next stage is the automatic pattern recognition for the classification problem. Classifiers based on K-nearest neighbours (K-NN) \cite{Khao2013}, Bayesian algorithms \cite{Wang2005}, support vector machines (SVM) \cite{Axelberh2007}, and artificial neural networks (ANN) \cite{Reaz2007} have been used for PQ pattern classification.

Several researches in PQ disturbance classification have used different combinations of dictionaries and classifiers to improve the PQ classification accuracy for PQ disturbances \cite{Granados}. Gargoom et al. \cite{Gargoom2007} used a K-NN classifier with statistical features computed either from a WT, a ST or a Hilbert-Huang transform (HHT), obtaining a classification accuracy around $82 \sim 89\%$. In \cite{Naik2011} and \cite{Lee2003}, the authors employed a ST to identify short durations disturbances, with an ANN classifier, obtaining a performance between $80 \sim 90\%$ for synthetic generated signals with 30-dB noise level. Similarly, Jayasree et al. used the energy of intrinsic mode functions (IMF) components computed from a HHT as features for an ANN classifier \cite{Jayasree2011}, obtaining $95 \sim 98\%$. Most recently, in \cite{Eristi2012, Verna2012, Reaz2007}, the authors have used the detailed coefficients from a WT, or some of its variant such as Wavelet multi-resolution analysis, together with  classifiers based on either a SVM or an ANN. They obtained a performance between $95 \sim 98\%$. 

To the best of our knowledge, previous works in PQ disturbance representation and classification have been limited to the use of single dictionaries for feature extraction step. This is, they only use either GT, HHT, WT or ST together with some classifiers \cite{Sharma2013, Vega2007, Eristi2012, Naik2011, Huang2009}. However, research in signal processing has shown that combining several dictionaries for signal analysis can improve the robustness and numerical stability in the feature extraction step \cite{kovacevic2008introduction, Chen1998, Lewicki98learningovercomplete}. Combinations of complete dictionaries for signal representations are also known as overcomplete representations (OR), and they are useful a) for increasing the richness of the representation by removing uncertainty when choosing the ``proper'' dictionary  \cite{kovacevic2008introduction}, b) for allowing a non-unique representation with the possibility of adaptation \cite{Chen1998}, c) for increasing the robustness in presence of noise, and d) for increasing the flexibility for matching any structure present in the data \cite{Lewicki98learningovercomplete}. However, OR increase the length of the coefficient vector of representation extracted from the signals, and may include information that is not relevant or that may be confusing for the classification step.

In order to remove redundant information, before the classification step, research communities in statistical signal processing, statistics and machine learning have proposed different methods for automatically performing basis selection in linear models. These methods are known as $\ell_1$ regularization methods or sparse linear models (SLM) \cite{bishop2007}. SLM highlight the relevant features in signal representation, making zero (or approximately zero) the contribution of less relevant ones. This property is known as sparsity \cite{murphy2012}. Taking advantage of the sparsity due to SLM, it is possible to build OR avoiding the redundant information among the dictionaries. This approach tends to increase the PQ representation accuracy due to OR, and reduces PQ classification complexity due to sparse tendency. Since our OR approach for PQ disturbances requires the analysis of variables (coefficient of representations) that are grouped per each dictionary (e.g. GT, WT or ST), in this  paper we employ a SLM known as Group Lasso for PQ disturbance classification. We evaluate the performance of Group Lasso using different classifiers with either one dictionary at a time or with OR. We employ the discrete-time form of the GT, WT and ST dictionaries. Finally, for the PQ classification step, we use different statistical classifiers, namely, K-NN, SVM, ANN, and two Bayesian classifiers based on linear (LDC) and quadratic (QDC) discriminant functions.

This paper is organized as follows. Materials and methods are described in section \ref{sec:methods}. In section \ref{sec:results&discussion}, we compare and discuss the results. Finally, section \ref{sec:conclusion} shows the conclusions.

\section{Materials and Methods}
\label{sec:methods}

\subsection{Power quality disturbances}
\label{subsec:pq}

PQ determines the fitness of electric power to consumer devices, evaluating the quality of voltage and current signals. We will use two concepts to describe PQ disturbances, namely, events and variations. Events are sudden distortions which occur in specific time intervals. On the other hand, variations are steady-state or quasi-steady-state disturbances which require continuous measurements \cite{Bollen2009}. This paper focusses on four types of disturbances: voltage or current variations (e.g. swell, sag and flicker), harmonic distortions, notching and transient events (e.g. impulsive and oscillatory).

\vspace{-10pt}
\subsubsection{Swell and sag events}
\vspace{-5pt}
swells describe an increase between $1.10$ and $1.80$ p.u. (per unit) over the RMS (Root Mean Square) value in an electric signal during an interval between $10$ ms and $1$ minute (short time), or over $1$ minute (long time) \cite{Bollen2006}. Opposite to swells, sags describe a decrease between $0.10$ and $0.90$ p.u. below the RMS value. This kind of distortions are commonly produced by tap charger, electrical loads with high values and electrical faults on the network \cite{kusko2007}.

\vspace{-10pt}
\subsubsection{Flicker variations}
\vspace{-5pt}
fluctuations between $\pm0.1\%$ and $\pm7\%$ respect to the RMS value with a frequency lower than $25$ Hz. Cycloconverters, electric arc furnace and high current starters are the main sources of flicker effects \cite{kusko2007, Pasko2012}.

\vspace{-10pt}
\subsubsection{Harmonic variations}
\vspace{-5pt}
distortions which present different frequency components in electric signals. These components correspond to integer multiples of the fundamental frequency \cite{Surajit2011}. Harmonics appear usually by the connection of non-linear devices to the network \cite{Pasko2012}.

\vspace{-10pt}
\subsubsection{Notching variations}
\vspace{-5pt}
they introduce high harmonic and non-harmonic frequencies in higher voltage systems, due to the normal operation of electronic devices when current is commutated from one phase to another \cite{kusko2007, Farazmand2014}. 

\vspace{-10pt}
\subsubsection{Transient events}
\vspace{-5pt}
they can be classified in two types, impulsive and oscillatory events. The impulsive transients are unidirectional events during a short time lapse between $50$ ns and $1$ ms. They are produced by electric discharges or inductive charge commutation (motors) \cite{kusko2007, Pasko2012}. Oscillatory transients are bidirectional events during an interval time between $0.3$ ms and $50$ ms. The oscillations occur commonly by power factor supplies using capacitors switching banks \cite{kusko2007, Pasko2012}.

\begin{figure*}[t!]
	\vspace{20pt}
	\centering
	\subfigure{\def\svgwidth{100pt} 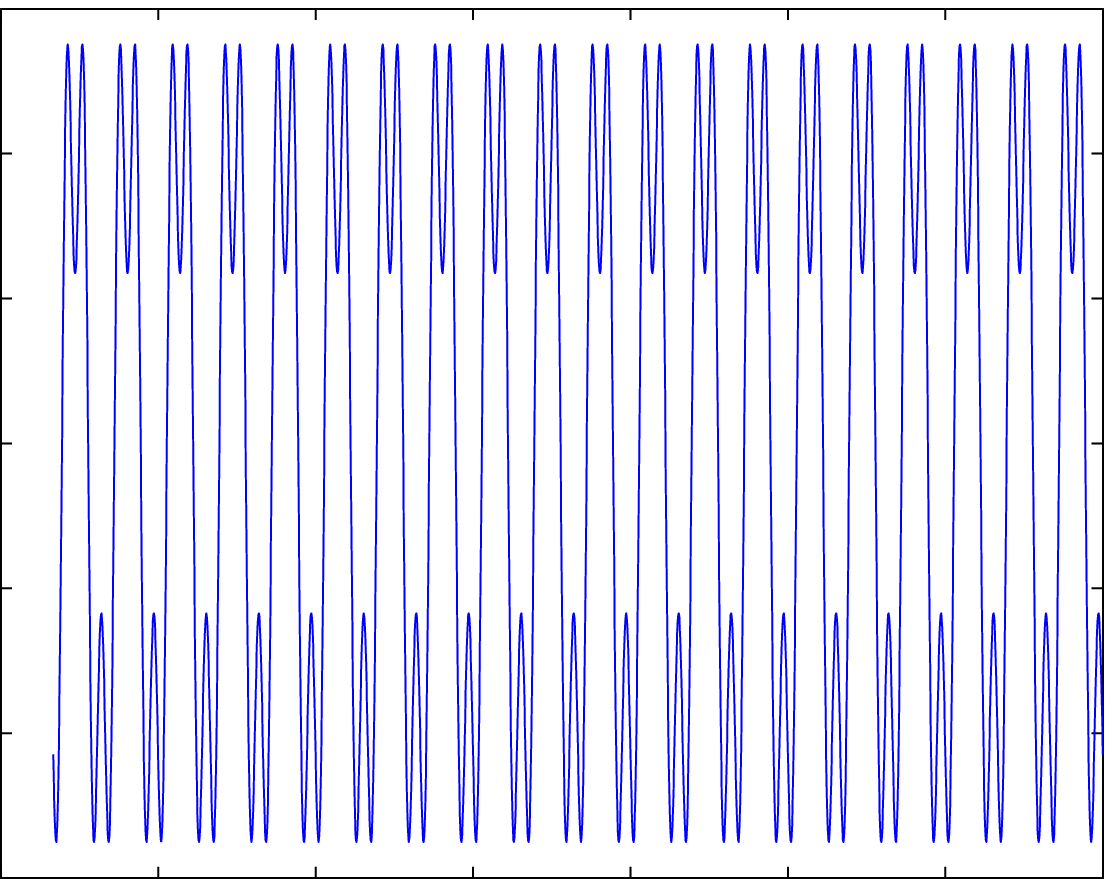}
	\hspace{15pt}
	\subfigure{\def\svgwidth{100pt} 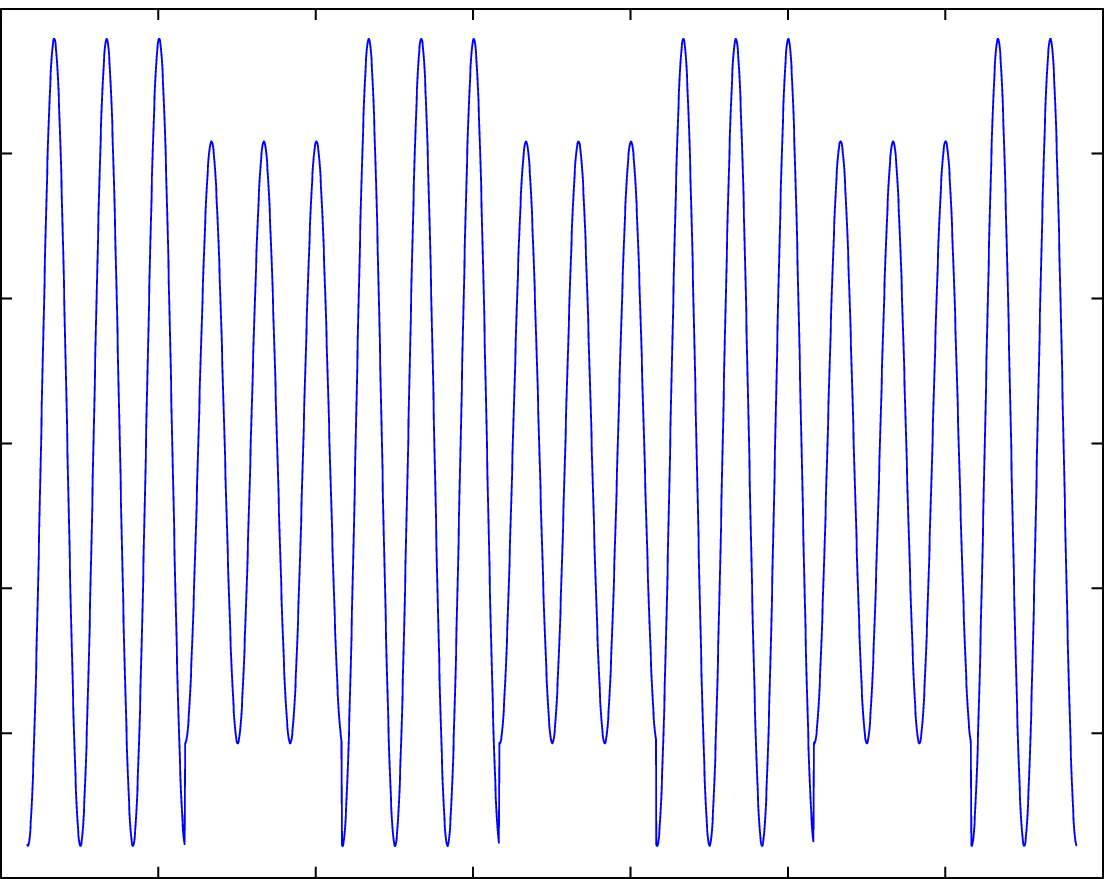}
	\hspace{15pt}
	\subfigure{\def\svgwidth{100pt} 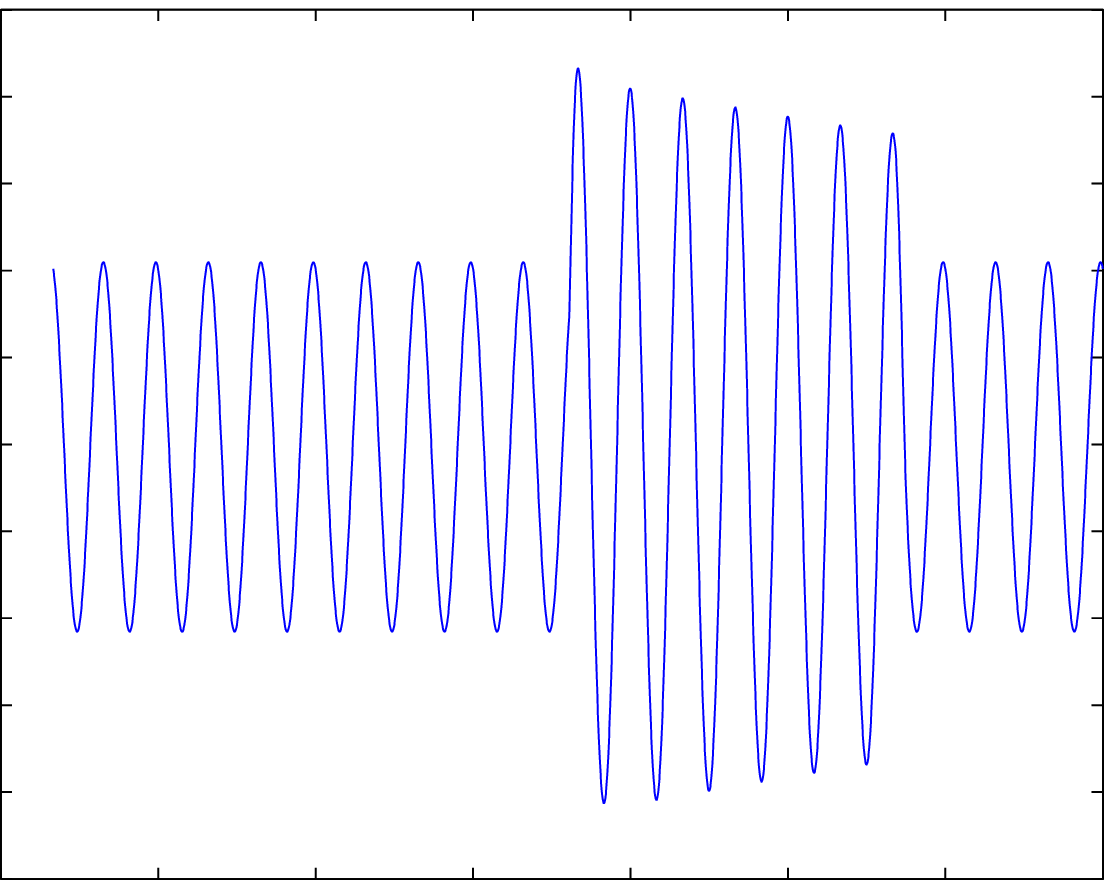}
	\hspace{15pt}
	\subfigure{\def\svgwidth{100pt} 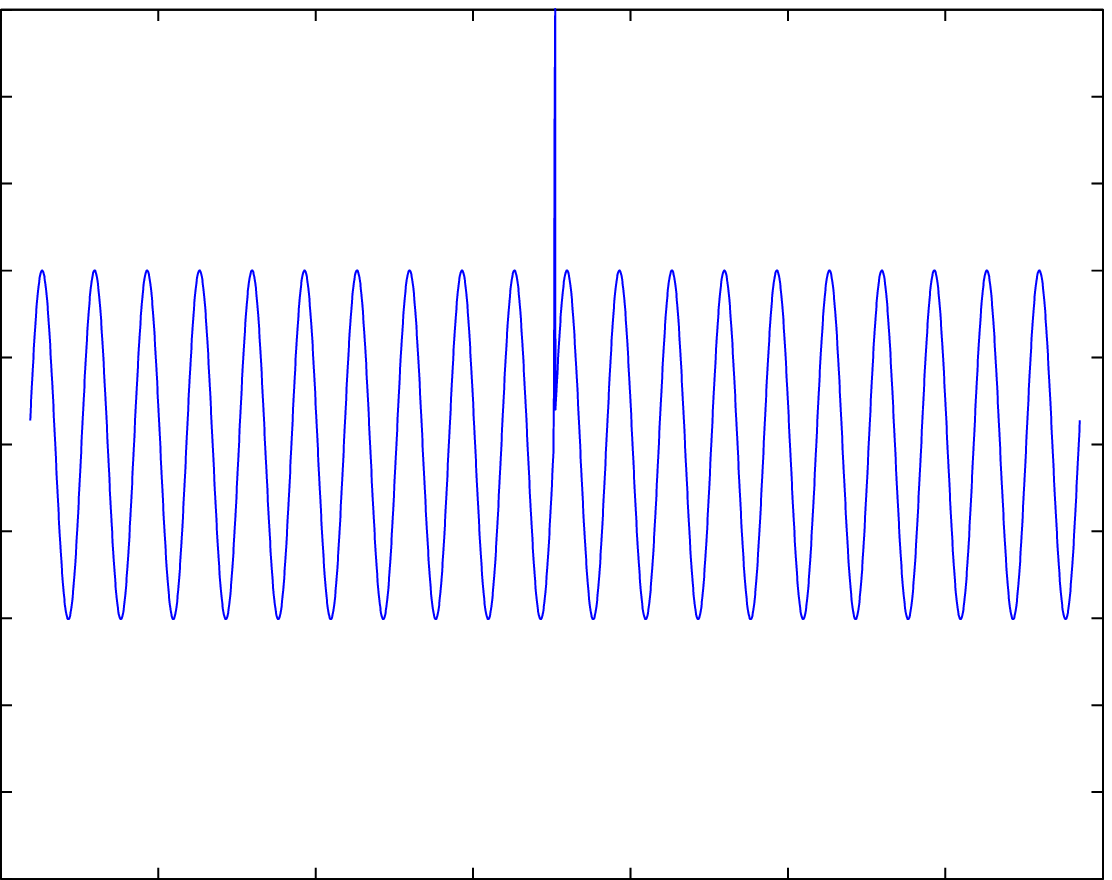}	
	
	\caption{PQ disturbance examples. Figure shows different PQ disturbances such as harmonic distortion, flicker variation, swell event, and impulsive transient (from left to right). The examples correspond to the synthetic dataset described in section \ref{subsec:procedure}}
	\label{fig:PQexample}
	\vspace{-5pt}
\end{figure*}

Table \ref{tab:PQissues} summarizes the characteristics of the PQ disturbances listed above. Figure \ref{fig:PQexample} shows examples of PQ distortions from the synthetic dataset described in section \ref{subsec:procedure}.

\begin{table}[h!]
	\caption{Summary of PQ disturbances.}
	{\scriptsize
	\begin{minipage}{\columnwidth}
		\centering
		\label{tab:PQissues}%
		\begin{tabular}{lccc}
			\toprule
			Disturbance & Spectral Content & Duration & Magnitude [p.u.]\\
			\midrule
			Events & & &\\
			\ \ 1. Swells & & &\\
			\ \ \ 1.1. Short time &       & $ 10 $ ms-$ 1 $ min & $ 1.1 $-$ 1.8 $ \\
			\ \ \ 2.1. Long time &       & $> 1$ min & $ 1.1 $-$ 1.8 $ \\		
			\ \ 2. Sags & & &\\
			\ \ \ 1.2. Short time    &       & $ 10 $ ms-$ 1 $ min & $ 0.1 $-$ 0.9 $ \\
			\ \ \ 2.2. Long time    &       & $> 1$ min & $ 0.1 $-$ 0.9 $ \\
			\ \ 3. Transients & & & \\
			\ \ \ 3.1. Impulsive & $5$ ns rise& $<50$ ns & \\
			\ \ \ 3.2. Oscillatory & & &\\
			\ \ \ \ 3.2.1. Low frequency & $<5$ KHz & $0.3$-$50$ ms & $0$-$4$ \\
			\ \ \ \ 3.2.2. Mid frequency & $5$-$500$ KHz & $20$ ms & $0$-$8$ \\
			\ \ \ \ 3.2.3. High frequency & $0.5$-$5$ MHz & $5$ ms & $0$-$4$ \\
			Variations & & &\\
			\ \ 4. Harmonic Distortion & k=$2$-$40$ (th) & steady state & $<0.2$ \\
			\ \ 5. Flicker & $<25$ Hz & intermittent & $0.001$-$0.07$ \\
			\ \ 6. Notching &       & steady state &  \\
			\bottomrule
		\end{tabular}%
		\footnotetext{\scriptsize Spectral Content and Duration correspond to the typical frequencies and time lapses. Magnitude represents the variation in amplitude and it is given in per unit (p.u.).}
	\end{minipage}%
	}
\end{table}

\subsection{Overcomplete representation}
\label{subsec:overcomplete}
Be $\By \in \mathds{R}^{N}$ a vector which represents a discrete-time signal (e.g. a PQ disturbance) of length $N$. A signal representation using a superposition of atoms (basis functions) $\phi_\gamma (t)$ is known as a complete representation \cite{Mallat1993}. Here, the collection of parametrized atoms $\Phi = (\phi_\gamma: \gamma \in \Gamma)$ is known as a dictionary, where the parameter $\gamma$ has different interpretations according to the indexation of the variables: frequency, time-frequency or time-scale values. In general, it is possible to assume that $\textbf{y}$ can be expressed by the following linear matrix equation
\begin{equation}
\By = \Phi \Bbeta,
\label{eq:LinearMatrixEquation}
\end{equation}
where $\Bbeta \in \mathds{R}^M$ and $\Phi \in \mathds{R}^{N\times M}$ are the vector of coefficients and the dictionary related to the representation, respectively. An overcomplete representation (OR) is obtained by the combination of several dictionaries in order to represent completely the signal y, obtaining a new dictionary given by $\Psi = (\Phi_\upsilon: \upsilon \in \Upsilon)$ \cite{Chen1998}. In this paper, we analyse the PQ disturbances in terms of the vector of coefficients $\Bbeta$ (analysis step). These coefficients are subsequently used for feature extraction, to feed the PQ disturbance classifiers. We also use $\Bbeta$ to reconstruct or synthesize (synthesis step) the PQ disturbances vector computed from the analysis step \cite{Chen1998}. We evaluate the quality of the reconstruction provided by the sparse vector obtained after applying Group Lasso.

\subsection{Sparse linear models for grouped variable selection: Group Lasso}
\label{subsec:sparse}
Tibshirani \cite{tibshirani1996} proposes a method for estimation in linear models known as Lasso (least absolute shrinkage and selection operator). To solve the inverse problem from the Equation \ref{eq:LinearMatrixEquation}, he demonstrated that Lasso obtains a sparse representation due to the minimization of the cost function given by
\begin{equation*}
\hat{\Bbeta}_\lambda =  \underset{\Bbeta} {\mbox{arg min}}    \bigg\{ \frac{1}{2}\|\By - \Phi\Bbeta\|_{2}^{2} + \lambda \|\Bbeta \|_1 \bigg\},
\label{eq:sp1}
\end{equation*}
where $\hat{\Bbeta}_\lambda \in \mathds{R}^M$ is the vector of estimated coefficients which depends of the regularization parameter $\lambda$, $\| \cdot \|_{2}$ and $\| \cdot \|_{1}$ correspond to the $\ell_2$-norm, and $\ell_1$-norm, respectively. Here, Lasso assumes that there is a direct and unique correspondence between parameters and variables (1:1 correspondence), performing the variable selection individually. However, the individual selection of variables produces a not satisfactory solution for grouped variables (e.g. coefficients $\Bbeta_\upsilon$ related to bases $\Phi_\upsilon$ from the dictionary $\Psi$) \cite{Meier2008}. To deal with grouped variable estimation, an extension of Lasso was proposed in \cite{Yuan2006}. This method is known as Group Lasso. Group Lasso assumes that the vector $\Bbeta$ is partitioned in $G$ groups where the penalty is an intermediate between $\ell_1$ and $\ell_2$ regularizations. Meier et al. \cite{Meier2008} showed that Group Lasso has the attractive property it performs the variable selection at the group level, promoting sparsity over $\hat{\Bbeta}_\lambda$ for large values of $\lambda$. For linear regression, the cost function for Group Lasso is given by
\begin{equation} 
\hat{\Bbeta}_\lambda = \underset{\Bbeta}
{\mbox{arg min}} \bigg\{\frac{1}{2}\|\By - \Phi\Bbeta\|_{2}^{2} +
\lambda \sum_{g=1}^{G}\|\Bbeta_g \|_{\textbf{K}_g} \bigg\},
\label{eq:sp2}
\end{equation}
with $\|\Bbeta_g \|_{\textbf{K}_g} = (\Bbeta_g^{\top} \textbf{K}_g \Bbeta_g)^{1/2}$. Yuag et al. \cite{Yuan2006} implemented a Group Lasso approach of the shooting algorithm proposed in \cite{Fu2006Lasso}, assuming $\textbf{K}_g = p_g I_{p_g}$ with $I_{p_g}$ an identity matrix which depends on the length of $g$-th group. According to this proposition, the necessary and sufficient condition for $\Bbeta = [\Bbeta_1^{\top}, \cdots, \Bbeta_G^{\top}]^{\top}$ to be a solution of Equation \eqref{eq:sp2} is given by
\begin{align}
\Bbeta_g = \bigg[ 1 - \frac{\lambda \sqrt{p_g}}{\|\BS_g\|} \bigg] \BS_g,
\label{eq:sp3}
\end{align}
where $\BS_g = \BPhi_g^{\top} (\By - \BPhi \Bbeta_{-g})$, with $\Bbeta_{-g} = [\Bbeta_{1}^{\top}, \cdots, \Bbeta_{g-1}^{\top}, \boldsymbol{0}, \Bbeta_{g+1}^{\top}, \cdots, \Bbeta_{G}^{\top}]^{\top}$. The solution of the Equation \eqref{eq:sp2} can therefore be obtained by iteratively applying the Equation \eqref{eq:sp3} for $g = 1, 2, \cdots, G$ \cite{Yuan2006}. In this paper, each group corresponds to a single dictionary, and $\Bbeta_g$ are the representation coefficients related to the $g$-th dictionary.

\subsection{Time-frequency and time-scale dictionaries}
\label{subsec:dictionaries}

\subsubsection{Short Time Fourier Transform (STFT)}
STFT is the general expression of time-frequency dictionaries. The STFT uses the concept of a window function to compute the Fourier transform in a specific interval of the analysis signal. The STFT of a discrete-time series $y[n]$ is giving by the expression
\begin{align*}
\operatorname{STFT}\{y[n]\} = \sum\limits_{n=-\infty}^{\infty} {y[n] h[n-m_l] \exp\{-j\omega_k [n-m_l]\} },
\end{align*}
where $h[n-m_l]$ is the window function with location $m_l$, and $f_k = \omega_k / 2\pi$ are the frequency terms which we use for the representation. In PQ disturbance, the Gabor transform (GT) is the most common STFT where $h[\cdot]$ is defined by a Gaussian function \cite{Sharma2013}. Other types of window functions used in PQ analysis are the Hamming function and the Hann function \cite{Sharma2013}. For real-valued representations, the atoms $\phi_{\gamma}[n]$ for GT dictionary follow
\begin{align*}
\frac{1}{\sqrt{2\pi \sigma^2}}\exp \left\{ -(n- m_l)^2/2\sigma^2 \right\}\cdot \cos \left( 2\pi f_k(n- m_l ) \right),
\end{align*}
where $\gamma = (f_k,m_l)$, $f_k = kf_o$ ($k = 1, 2, \cdots, K$), and $m_{l} = l m_o$ ($l = 1, 2, \cdots, L$). Here, $K$ (total frequencies), $L$ (total locations), $f_o$ (fundamental frequency) and $m_o$ (fundamental location) are constants. The main disadvantage of GT in PQ disturbances is related to its resolution for signal representation. A higher resolution in time leads to a lower resolution in frequency, and vice versa \cite{Sharma2013}. In addition, we need to fix the length scale $\sigma$ \cite{Vega2007}. 

\subsubsection{Wavelet Transform (WT)}
WT is other example of time-frequency transforms applied to PQ disturbances \cite{Eristi2012}. Similarly to STFT, the WT uses a time-scale dictionary characterized by irregular functions called wavelets. Contrary to STFT, WT gives the possibility to vary the scale factor, obtaining higher resolutions in time when it is required. Here, we can fix the length of the wavelet to detect high frequencies \cite{Vega2007}. WT is described by
\begin{align*}
\mbox{CWT} \{y[n]\} = \frac{1}{\sqrt{\sigma_\nu}} \sum\limits_{n=-\infty}^{\infty} {y[n] \psi \left[ \frac{n-m_l}{\sigma_\nu} \right] },
\end{align*}	
where $\psi[(n-m_l)/\sigma_\nu]$ is the wavelet function with locations $m_l$ and length-scale $\sigma_\nu$. Mexican Hat Wavelet Transform (MHWT), Daubechies, Morlets and Symlets wavelets have been applied to PQ disturbance representation \cite{Eristi2012}. In this paper, we use the MHWT. Its atoms $\phi_\gamma[n]$ can be expressed by the following form
\begin{align*}
\frac{2}{\sqrt{3\sigma_\nu} \;\pi^{1/4}} [1-(n-m_l)^2/\sigma_\nu^2] \cdot \exp \{-(n-m_l)^2/2\sigma_\nu^2\},
\end{align*}
where $\gamma = (m_l,\sigma_\nu)$,  $\sigma_\nu = \nu\sigma_o$ ($\nu = 1, 2, \cdots, V$), and $m_{l} = l m_o$. Here, $V$ (total scales), $L$, $\sigma_o$ (principal scale) and $m_o$ are constants. The WT has been used to characterize PQ disturbances such as voltage variation (e.g. swells and sags) \cite{Eristi2012}, and transients distortions (e.g. impulsive and oscillatory), due to its variable length-scale \cite{Vega2007}. 
\subsubsection{Stockwell Transform (ST)}
ST is a generalization of the STFT, extending the continuous WT and overcoming some of its disadvantages. ST includes a correction factor which gives information about phase contribution in the frequency domain, improving the performance for PQ representations. The phase correction can characterize the real and the imaginary spectrum components independently \cite{Naik2011}. The ST combines the time-frequency and time-scale concepts, given
\begin{align*}
\mbox{ST} \{y[n]\} =  \sum\limits_{n=-\infty}^{\infty} {y[n] g\left[n-m_l,\sigma_\nu \right] \exp \{-j \omega_k (n-m_l) \} } ,
\end{align*}
where $g[n - m_l,\sigma]$ is a Gaussian window with location $m_l$ and length-scale $\sigma_\nu$. The atoms $\phi_\gamma[n]$ for ST are given by
\begin{align*}
\frac{1}{\sqrt{2\pi \sigma_\nu^2}}\exp \left\{ -(n-m_l)^2/2\sigma_\nu^2 \right\}\cdot \cos \left( 2\pi f_k(n-m_l) \right),
\end{align*}
where $\gamma = (f_k,m_l,\sigma_\nu)$, $f_k = kf_o$, $m_{l} = l m_o$, and $\sigma_\nu = \nu\sigma_o$. ST shows a high performance in PQ representations for several types of distortions such as swell, sag and flickers \cite{Naik2011, Huang2009}.

In general, it is possible to express many dictionaries in a matrix form $\Phi = (\phi_\gamma: \gamma \in \Gamma)$. Also, we can combine several dictionaries to achieve a new and most robust dictionary by $\Psi = (\Phi_\upsilon: \upsilon \in \Upsilon)$, where $\gamma$ and $\upsilon$ are indexation parameters. It is noteworthy that these dictionaries do not need to be orthogonal to obtain an OR \cite{Chen1998}.

\begin{table*}
	\caption{Parametric equation for PQ disturbance simulations.} 
	{\scriptsize
	\begin{minipage}{\textwidth}
		\centering
		\label{tab:dataset2}
		\begin{tabular}{cccc}
			\toprule
			PQ disturbance & Notation & Controlling Parameters & Parametric Equation \\ \midrule
			\multirow{2}{*}{Pure Signal} & \multirow{2}{*}{$-$} & \multirow{2}{*}{$-$} & \multirow{2}{*}{$y(t) = \sin(\omega t)$}  \\
			&&&\\ 
			Harmonic & \multirow{2}{*}{$C_1$} & \multirow{2}{*}{$0.05 \le \alpha_k \le 0.15, \; \sum_{k} \alpha_k^2 = 1$} & \multirow{2}{*}{$y(t) = \sum_{k} \alpha_k \sin(k \omega t)$ con $k = 3, 5, 7, 9.$}\\ 
			Distortion & & & \\
			\multirow{2}{*}{Swell} & \multirow{2}{*}{$C_2$} & \multirow{2}{*}{$0.1 \le \alpha \le 0.8,\; 6 \le t_2-t_1 \le 10T$} & \multirow{2}{*}{$y(t) = [1 + \alpha\{u(t_2)-u(t_1)\}]\sin(\omega t)$}\\
			&&& \\
			\multirow{2}{*}{Sag} & \multirow{2}{*}{$C_3$} & \multirow{2}{*}{$0.1 \le \alpha \le 0.8,\; 6 \le t_2-t_1 \le 10T$} & \multirow{2}{*}{$y(t) = [1 - \alpha\{u(t_2)-u(t_1)\}]\sin(\omega t)$} \\
			& & & \\
			Flicker & \multirow{2}{*}{$C_4$} & \multirow{2}{*}{$0.1 \le \alpha \le 0.2, \; 5 \le f_f \le 25$Hz}  & \multirow{2}{*}{$y(t) = [1+\alpha \sin(2\pi f_{f} t)] \sin(\omega t)$} \\
			Effect & & & \\
			\multirow{2}{*}{Notch} & \multirow{2}{*}{$C_5$} & \multirow{2}{*}{$0.1 \le \alpha_k \le 0.4, \; 0.01T \le \Delta t_k \le 0.05T$}  & \multirow{2}{*}{$y(t) = \sin(\omega t) - \mbox{sign}\{\sin(\omega t)\} \sum_{k}\alpha_k [u(t_k + \Delta t_k/2)-u(t_k - \Delta t_k/2)]$} \\
			& & & \\
			Impulsive & \multirow{2}{*}{$C_6$} & \multirow{2}{*}{$1.5 \le \alpha \le 2.5, \; 0.01T\le t_2 - t_1 \le 0.02T $} & \multirow{2}{*}{$y(t) = \sin(\omega t) + \alpha \ \mbox{sign}\{\sin(\omega t)\} [u(t_2)-u(t_1)]$} \\ 
			Transient & & & \\
			& & & \\ 
			Oscillatory & \multirow{2}{*}{$C_7$} & $0.1 \le \alpha \le 0.9, \; 0.01T\le t_2 - t_1 \le 0.03T$ & \multirow{2}{*}{$y(t) = \sin(\omega t) + \alpha \exp\{-t/\beta\} \sin(k\omega t) [u(t_2)-u(t_1)]$} \\
			Transient & & $0.1 \le \beta \le 0.2, \; 5 \le k \le 80$ & \\
			\bottomrule
		\end{tabular}
		\footnotetext{\scriptsize $T$ and $\omega$ are the period and the angular frequency, respectively. $u\{ \cdot \}$ and sign$\{ \cdot \}$ are the unit step and the sign functions, respectively.}
	\end{minipage}
	}
\end{table*}

\subsection{Classifiers for automatic recognition}
\label{subsec:classiffier}
PQ disturbance classification can be seen as a pattern recognition problem, in which each class corresponds to a type of PQ disturbance. We start with a set $\{\Bx_p,\ell_p\}_{p=1}^{P}$, where $\Bx_p \in \mathds{R}^{\mathcal{D}}$ is a feature vector extracted from the signal $p$ (we will see later how the elements in $\Bx_p$ are computed from the vector of coefficients $\Bbeta_p$), and $\ell_p$ is the corresponding label or class for signal $p$. $P$ is the total number of samples. The automatic recognition of disturbances is made of two stages: training and testing stages. In the first stage, the classifiers use a set of samples $\{\Bx_{t}, \ell_{t} \}_{t=1}^{T}$, to learn a mapping between the input space and the different PQ disturbances. In the second stage, the classifiers use the information obtained from the training set to predict the labels of a new set of samples $\{\Bx_{u} \}_{u=1}^{U}$. Notice that $T$ and $U$ correspond to the number of samples for the training and test datasets, respectively. In this subsection, we give a brief description of the classifiers we used to validate the performance of SLM in the PQ classification step. The classifiers that we use are state-of-the-art, and the theory behind them can be found in deep in any pattern recognition or machine learning textbook \cite{bishop2007, murphy2012}.  

\vspace{-10pt}
\subsubsection{K-nearest neighbours (K-NN)}
\vspace{-5pt}
K-NN is a non-parametric method where the labels of the K nearest points in the training set are taken  into account to predict the labels for the test set \cite{Khao2013}. Here, the proximity between neighbours is defined by the Euclidean distance \cite{bishop2007}.

\vspace{-10pt}
\subsubsection{Bayesian classifiers}
\vspace{-5pt}
there are two types of non-parametric Bayesian classifiers commonly used for patter recognition, namely, the linear (LDC) and the quadratic (QDC) discriminant classifiers. LDC finds a linear combination of features which is used as a linear classifier. This method assumes normal densities with equal covariance for each class, and the joint covariance is given by the weighted average of the class covariances. QDC also assumes that measurements for each class are normally distributed, but there is not the assumption that covariances for each class are identical. Both methods consist of the computation of the sample means and the sample covariances of each class \cite{bishop2007,murphy2012}.

\vspace{-10pt}
\subsubsection{Support vector machines (SVM)}
\vspace{-5pt}
SVM are decision machines that maximize the separating margin between two classes given a set of training data \cite{murphy2012}. The risk function of SVM involves the use of Lagrange multipliers, a set of Karush-Kuhn-Tucker conditions, and variables to control the trade-off between the slack variable penalty and a boundary that separate both classes \cite{bishop2007}. The SVM maximize the Lagrangian dual risk function which involves inner products between training data, which later are replaced for kernel functions. Some common kernels for PQ disturbance classification are the exponential, and the Gaussian kernels (or radial basis function, RBF) \cite{Axelberh2007} \cite{Lobos2006}.

\vspace{-10pt}
\subsubsection{Artificial neural networks (ANN)}
\vspace{-5pt}
ANN use non-linear activation functions over the outputs of linear models which can be adjusted during the training stage \cite{bishop2007}. More details can be found in \cite{bishop2007}. The most common classifiers for PQ disturbance are based on ANN and the feed-forward back-propagation scheme \cite{Reaz2007, Naik2011}.

\subsection{Procedure}
\label{subsec:procedure}

\vspace{-10pt}
\subsubsection{Dataset}
\vspace{-5pt}
Table \ref{tab:dataset2} shows the parametric equations used to construct the synthetic dataset for different types of PQ disturbances, namely, swells, sags, flicker effects, notches, impulsive and oscillatory transients. Those equations are based on \cite{Lobos2006, Salama2004, Verna2012,Roy2012}. We generated $1330$ PQ disturbances with $190$ samples per each type. Each disturbance has $2101$ discrete-time values for an interval between $0$ and $0.7$ seconds.

\vspace{-10pt}
\subsubsection{Group Lasso}
\label{subsubsec:GroupLasso}
\vspace{-5pt}
we implement the shooting algorithm for Group Lasso described in section \ref{subsec:sparse}, using a penalty factor $\lambda = 1\times10^{-3}$, and a convergence tolerance equal to $1\times10^{-15}$. These values were selected manually in order to obtain a signal reconstruction error lower than $1 \times 10^{-3}$, and for preserving relevant representation coefficients. Here, each group corresponds to a single dictionary, and $\Bbeta_g$ are the representation coefficients related to the $g$-th dictionary. The percentage of sparsity for the coefficients $\Bbeta$ is computed by the sum of the number of coefficient lower than $1\times 10^{-4}$, and dividing the result by the length of the vector $\Bbeta$.

\vspace{-10pt}
\subsubsection{Dictionaries}
\vspace{-5pt}
for each time-frequency and time-scale dictionary, we use a location factor $m_o = 0.005s$. For the GT, we focus in the first $40$ harmonics with a length-scale $\sigma = 0.005$s. For MHWT, we used as scale factor $\sigma_o = 0.005s$ with $V=10$. For ST, we used the first $6$ harmonics, and $\sigma_o = 0.005s$ with $V=6$. However, we add a cosine/sine dictionary for all the cases with the first $40$ harmonics, aiming to improve the PQ synthesis results. We denote it as the Harmonics dictionary. Finally, in order evaluate the SLM performance using OR, we combine the GT, MHWT and ST dictionaries which we denote as GWST dictionary.

\vspace{-10pt}
\subsubsection{Feature extraction}
\vspace{-5pt}
\label{subsec:feature}
after applying the dictionaries proposed in subsection \ref{subsec:dictionaries}, we compute the following features from $\{\Bbeta_p\}_{p=1}^P$ to create the feature vectors $\{\mathbf{x}_p\}_{p=1}^P$ for each signal $p$: mean of the absolute values ($F_1$), standard deviation ($F_2$), kurtosis ($F_3$), Shannon's energy ($F_4$) and RMS value ($F_5$). We also include the mean of the absolute values of the derivative $\Bbeta'$ ($F_6$)\footnote{Let $\Bbeta = [\beta_1, \beta_2, \cdots, \beta_M ]^{\top}$, the features were calculated as follows
	
	$F_1 = \frac{1}{M} \sum_{i=1}^{M} |\beta_i|,\quad \ F_2 = \left[\frac{1}{M-1} \sum_{i=1}^{M} (\beta_i - \bar{\beta})^2 \right]^{1/2}, F_3 = { \left[\frac{1}{M} \sum_{i=1}^{M} (\beta_i-\bar{\beta})^4\right]}/{\left[ \frac{1}{M} \sum_{i=1}^{M} (\beta_i - \bar{\beta})^2 \right]^2}$, $F_4 = \frac{1}{2} \sum_{i=1}^{M} \beta_i^2$, 
	
	$F_5 = \sqrt{\frac{1}{M} \sum_{i=1}^{M} |\beta_i|^2}, \quad F_6 = \frac{1}{M} \sum_{i=1}^{M} |\beta'_i|$,
	
	where $\Bbeta'$ is the derivative of $\Bbeta$ and $\bar{\beta}
	= \frac{1}{M} \sum_{i=1}^{M} \beta_i$.  }.
For the PQ disturbance classification step, we normalize the features by subtracting their mean values, and dividing the result using their standard deviation.  

\vspace{-10pt}
\subsubsection{Classifiers}
\label{subsubsec:classifier}
\vspace{-5pt}
for the classifier based on ANN, we use the Neural Network Toolbox provided for Matlab R2013a. We use an ANN made of three hidden layers with $20$, $15$ and $10$ neurons in the first, second, and third layer, respectively, with sigmoid transfer functions \cite{Monedero2007}. We employ the Levenberg-Marquardt scheme using the MSE criteria for the training stage. For the other classifiers described in   \ref{subsec:classiffier}, we use the Matlab Toolbox for Pattern Recognition (PRTools Toolbox\footnote{PRTools Toolbox is available at \url{http://prtools.org/}.}). For the SVM, we  use a RBF kernel $k(\mathbf{x}, \mathbf{x}')= \exp(-\|\mathbf{x} -\mathbf{x}'\|^2/\sigma^2)$, where $\sigma$ is known as the bandwidth parameter. The bandwidth parameter, and the regularization parameter for the SVM are tuned by cross-validation. We test the classifiers twenty times with different training sets. We select randomly the $70\%$ of the total samples per each type of PQ disturbance to train the classifiers (133 samples per class), and then we use the other $30\%$ for test stage (57 samples per class). The performance for each test experiment is computed by the sum of the successful classifications of PQ disturbances, and dividing the result using the total number of test samples. Finally, we compute the mean and the standard deviation of the performance obtained in all the experiments.

\section{Results and Discussions}
\label{sec:results&discussion}

\begin{figure}[t!]
	\centering
	\vspace{20pt}
	\subfigure[\label{fig:S1}]{
		\def\svgwidth{90pt} 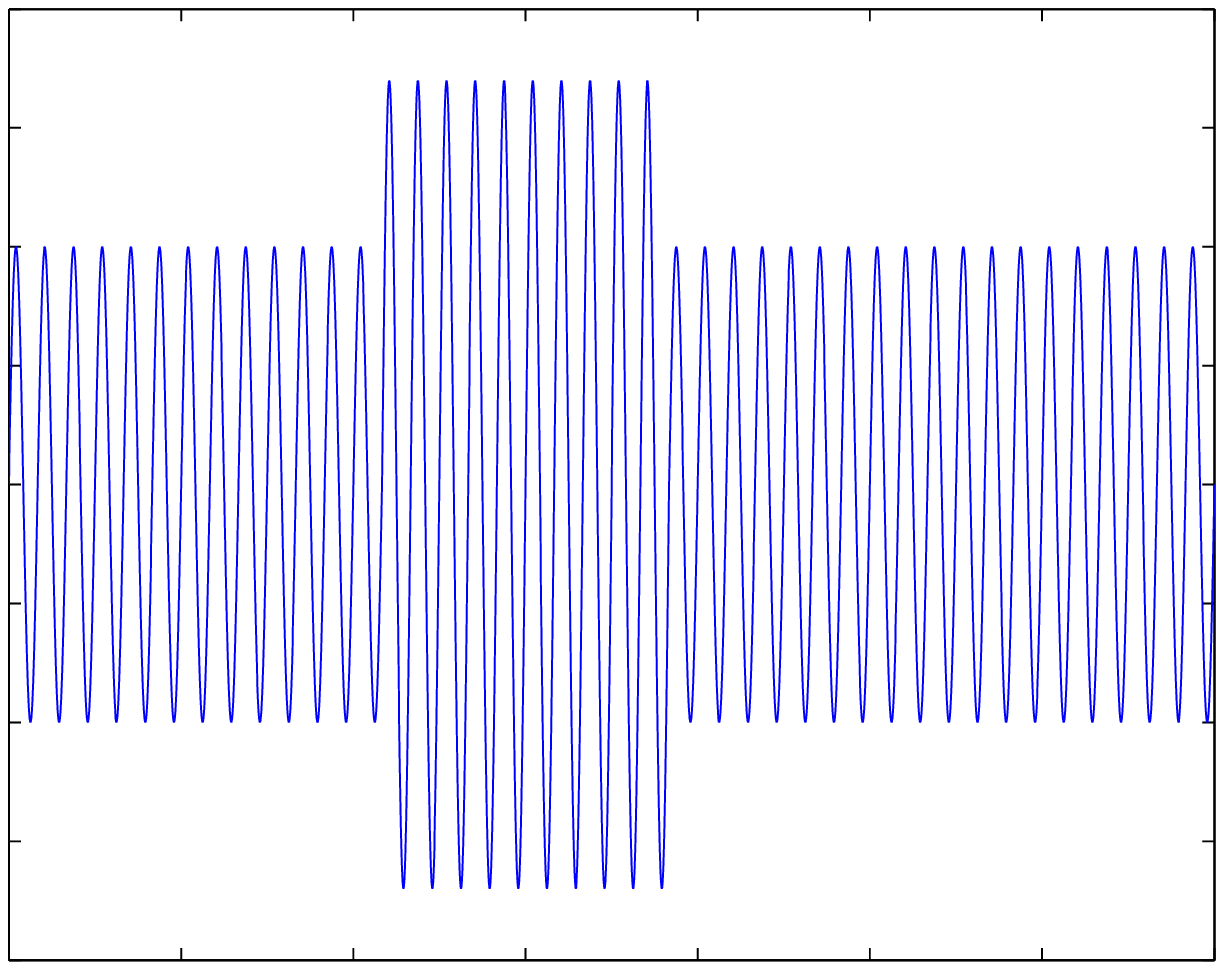 }
	\hspace{7pt}		
	\subfigure[\label{fig:S2}]{
		\def\svgwidth{90pt} 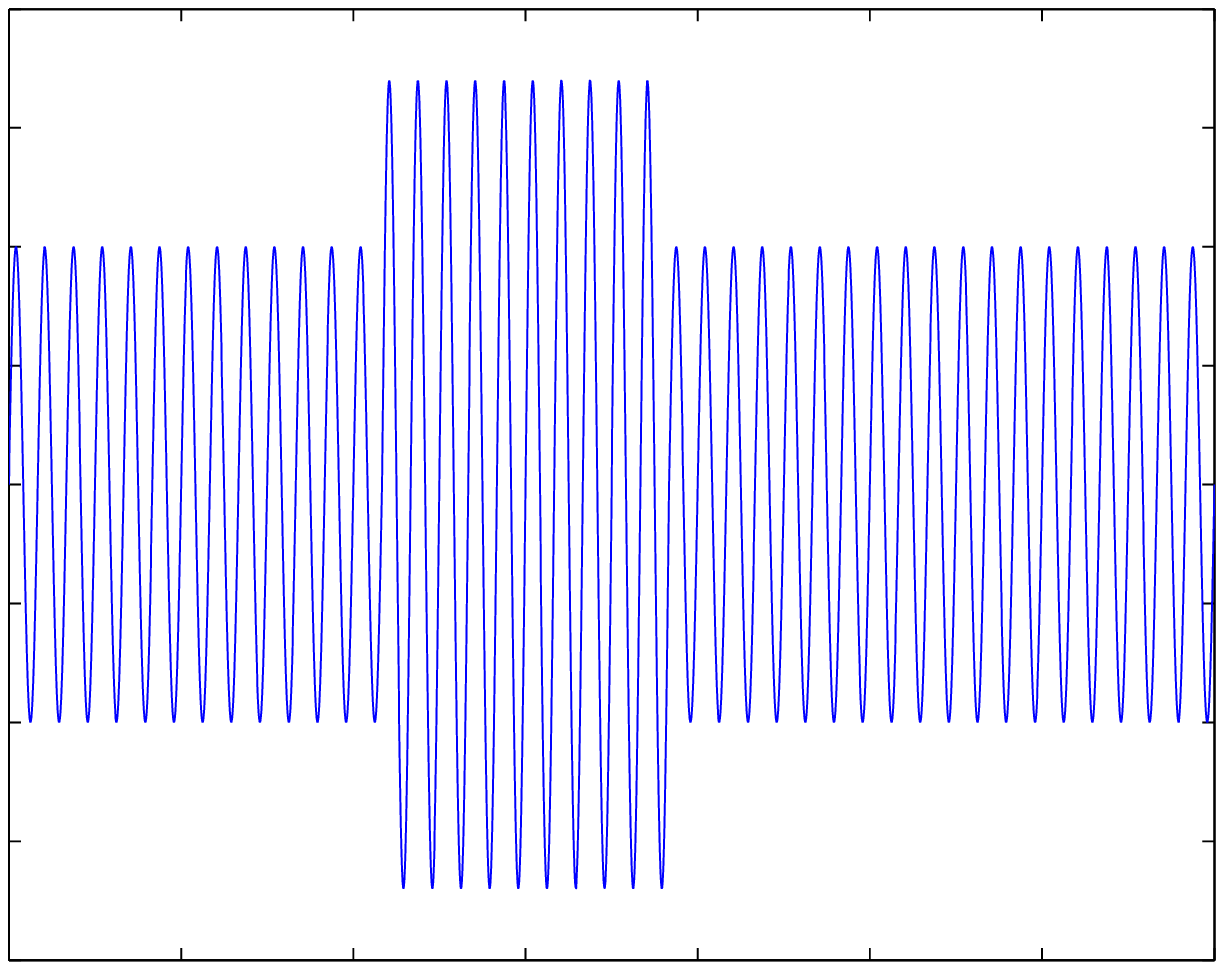 }
	\hspace{7pt}		
	\subfigure[\label{fig:S4}]{
		\def\svgwidth{90pt} 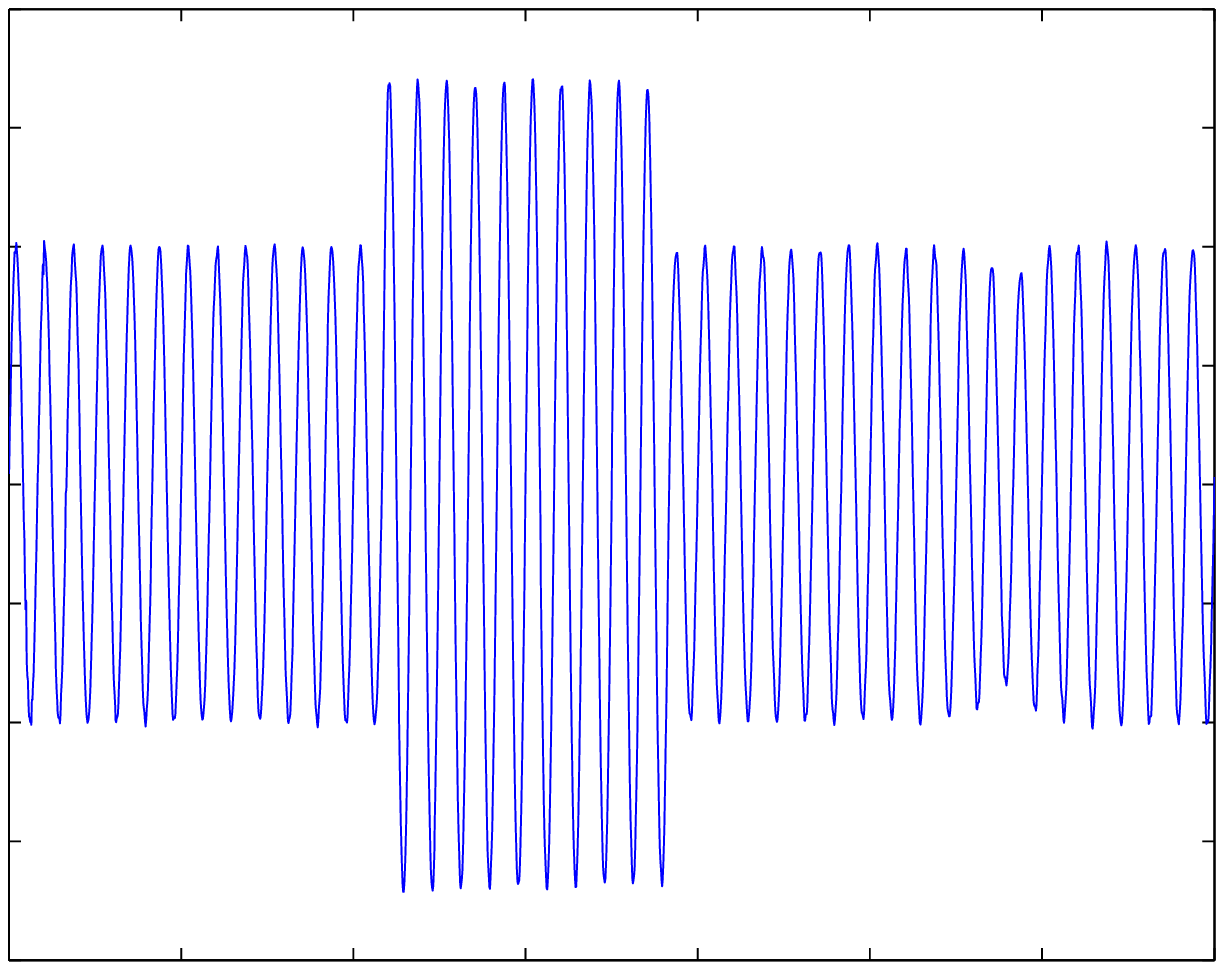 }\\ 
	\vspace{15pt}
	\subfigure[\label{fig:B1}]{
		\def\svgwidth{90pt} 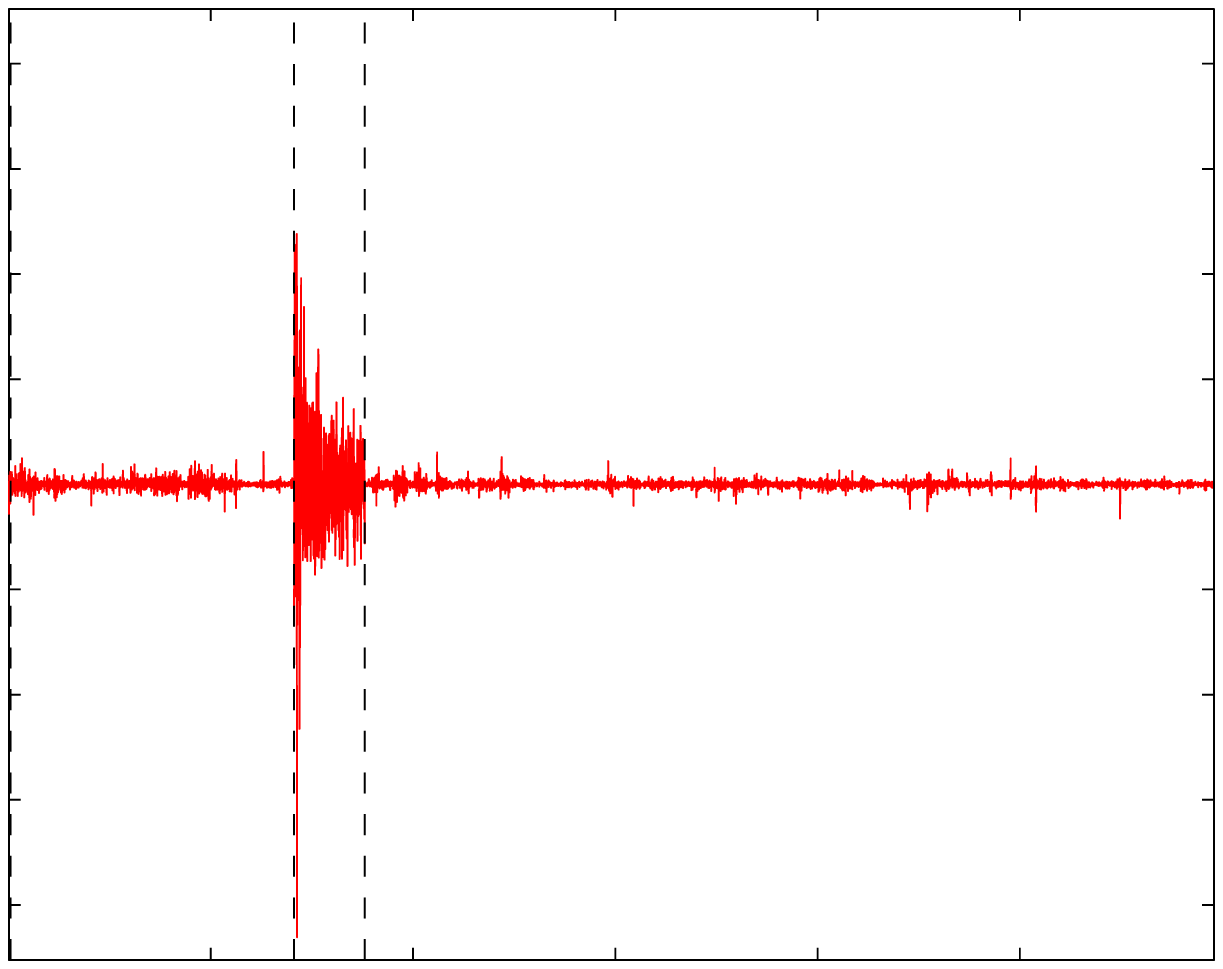 }
	\hspace{7pt}	
	\subfigure[\label{fig:B3}]{
		\def\svgwidth{90pt} 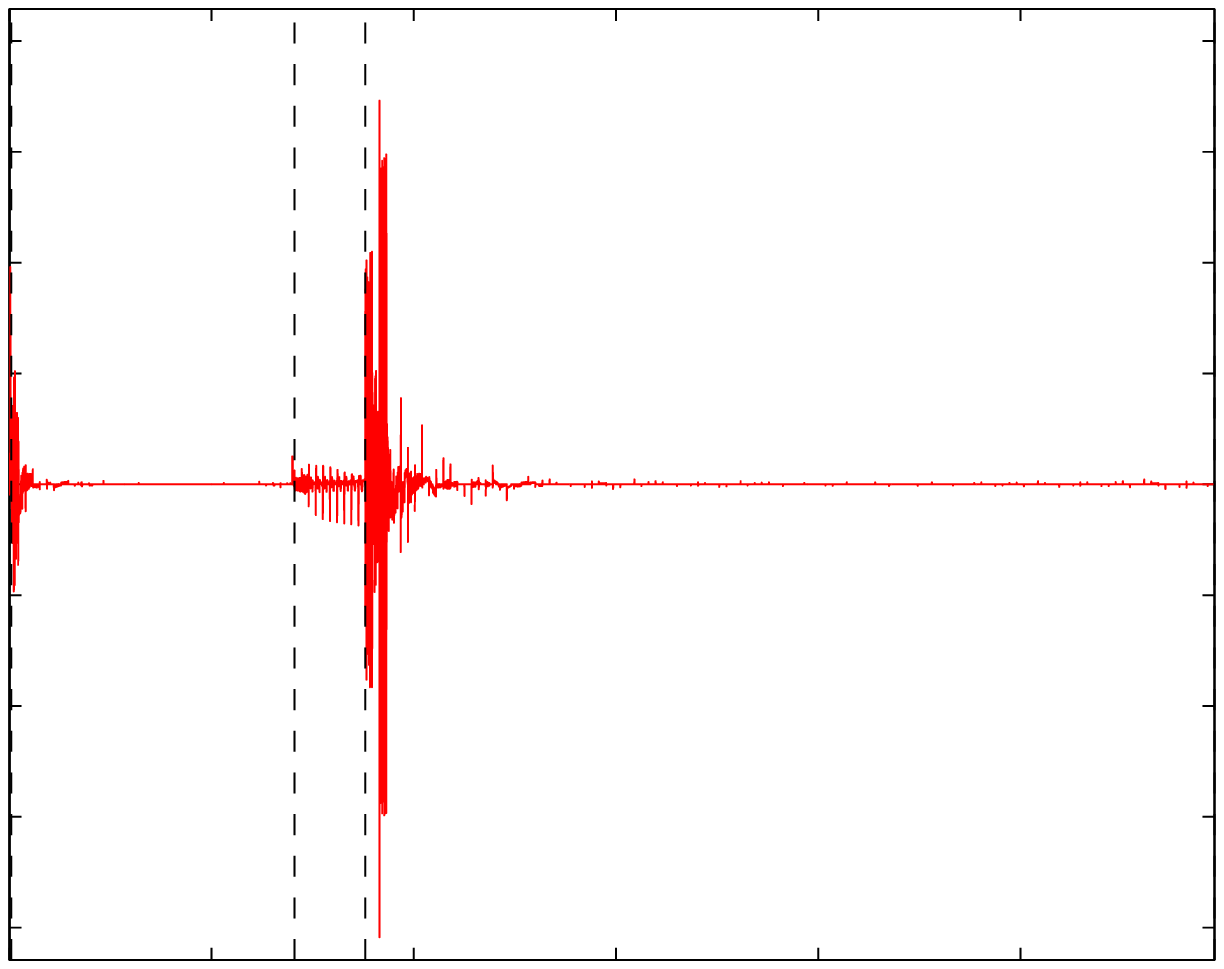 } 
	
	\caption{Synthesis and analysis step results for a swell event using GWST dictionary. In \subref{fig:S1} we show a swell example. The synthesis step without SLM and using Group Lasso are showed in \subref{fig:S2} and \subref{fig:S4}, respectively. Finally, we show the coefficient $\Bbeta$ used for synthesis step from Figures \subref{fig:S2} and \subref{fig:S4}, in Figures \subref{fig:B1} and \subref{fig:B3}.}
	\label{fig:S}
\end{figure}

To observe the advantages of sparse linear models (SLM) for PQ representation, Figure \ref{fig:S} shows the synthesis and the analysis steps for a single example of a swell event. The swell disturbance used as example appears in Figure \subref{fig:S1}. The result of the synthesis step is shown for two cases: in \subref{fig:S2} without sparsity, and in \subref{fig:S4} using Group Lasso. For this result, we use the GWST dictionary which we obtain by combining the GT, MHWT and ST. Both methods can synthesize the PQ distortion in Figure \subref{fig:S1}, ensuring a low reconstruction error. Moreover, we perform the analysis step in Figures \subref{fig:B1} and \subref{fig:B3} for the same example. The first, second and third interval separated by the dash line correspond to the coefficients obtained by GT, WT and ST, respectively. To represent the swell example, we see the method without Group Lasso in \subref{fig:B1} tends to use all the coefficients from $\Bbeta$. However, from Figure \subref{fig:B3}, Group Lasso achieves the synthesis of the PQ disturbance using slightly less than the $60\%$ of the total coefficients available, selecting a few number of coefficient from each dictionary. Finally, we see that the magnitude of the coefficients obtained with Group Lasso is far smaller compared to the magnitudes obtained without SLM due to the regularization parameter used in Group Lasso.

\begin{table}[t!]
	\begin{minipage}{\columnwidth}
		\centering
		\caption{Sparsity percentages using GWST without Group Lasso.}
		{\scriptsize
		\begin{minipage}{\textwidth}
			\centering		
			\begin{tabular}{ccccc}
				\toprule
				Type  & Harmonics & GT        &  MHWT     &  ST       \\
				\midrule			
				Harmonic 	& 0         &    0.0043 &         0 &    0.0150 \\
				Swell 		& 0         &    0.0034 &    0.0001 &    0.0054 \\
				Sag 		& 0         &    0.0045 &         0 &    0.0083 \\
				Flicker 	& 0         &    0.0041 &         0 &    0.0057 \\
				Notch 		& 0         &         0 &         0 &    0.0015 \\
				Impulsive 	& 0         &    0.0041 &         0 &    0.0087 \\
				Oscillatory & 0         &    0.0032 &         0 &    0.0048 \\
				\bottomrule
			\end{tabular}
		\end{minipage}		
		}
		\label{tab:sparse_GWST_none}
		\vspace{7pt}	
		
		\caption{Sparsity percentages using GWST with Group Lasso.}
		{\scriptsize		
		\begin{minipage}{\textwidth}
			\centering	
			\begin{tabular}{cccrc}
				\toprule
				Type  & Harmonics & GT        &  \multicolumn{1}{c}{MHWT} &  ST      \\
				\midrule	
				Harmonic 	& 38.1053   &   81.0351 &   10.6900 &   56.1971 \\
				Swell 		& 52.3421   &   91.0189 &    7.5048 &   64.1343 \\
				Sag 		& 51.7632   &   91.0485 &    7.6622 &   63.9687 \\
				Flicker 	& 61.6579   &   92.5543 &    9.5789 &   65.4090 \\
				Notch 		& 19.4737   &   51.0748 &   10.8880 &   32.6568 \\
				Impulsive 	& 39.9211   &   47.8662 &    8.0383 &   23.5953 \\
				Oscillatory & 39.4737   &   67.7954 &   10.3418 &   44.8384 \\
				\bottomrule
			\end{tabular}
			\footnotetext{\centering \scriptsize GT: Gabor transform. MHWT: Mexican hat Wavelet transform. ST: Stockwell transforms.}
		\end{minipage}	
		}
		\label{tab:sparse_GWST_GLasso}
	\end{minipage}
\end{table}

In order to quantify the level of sparsity produced by Group Lasso over the GWST representation, the synthesis step was performed over all the PQ disturbance examples from the synthetic dataset described in section \ref{subsec:procedure}. Tables \ref{tab:sparse_GWST_none} and \ref{tab:sparse_GWST_GLasso} show the sparsity percentages produced by the method without sparsity, and using Group Lasso, respectively. According to section \ref{subsubsec:GroupLasso}, we compute the sparsity percentages per each type of PQ disturbance (rows), and each dictionary (columns). Table \ref{tab:sparse_GWST_none} shows that methods without sparsity tend to use all the coefficient of representation from GWST, making more difficult subsequent PQ studies (e.g. PQ analysis and PQ disturbance classification). According to Table \ref{tab:sparse_GWST_GLasso}, Group Lasso automatically selects the most representative coefficients required to fully represent the PQ disturbances. Also, we notice that the Harmonics, the MHWT and the ST dictionaries contain more relevant components in terms of the disturbance representation, when compared to the GT dictionary.

\begin{table*}[t!]
	\caption{Performance of the classifiers without SLM and using Group Lasso for GT, MHWT, ST and combining all the dictionaries (GWST). The results by dictionaries and classifiers employed are indexed by rows and columns, respectively.}
	\label{tab:result}
	{\scriptsize			
	\begin{minipage}{\textwidth}
		\centering
		\begin{tabular}{rcccccc}
			\toprule
			\multicolumn{1}{c}{\multirow{2}{*}{Dictionary}} & 1-NN & 3-NN & LDC & QDC & SVM & ANN \\
			& $\mu \pm \sigma$ & $\mu \pm \sigma$ & $\mu \pm \sigma$ & $\mu \pm \sigma$ & $\mu \pm \sigma$ & $\mu \pm \sigma$ \\\hline
			\multicolumn{7}{|c|}{without SLM} \\ \hline
			GT + Harmonics & $0.7832  \pm  0.0129$  &  $0.7189  \pm  0.0156$  &  $0.4244 \pm  0.0175$  &  $0.6023  \pm  0.0165$  &  $0.7756  \pm  0.0165$  &  $0.6373  \pm  0.0489$ \\
			MHWT + Harmonics &  $0.7614  \pm  0.0161$  &  $0.7484  \pm  0.0183$  &  $0.5743  \pm  0.0229$  &  $0.6637  \pm  0.0175$  &  $0.7514  \pm  0.0166$  &  $0.7579  \pm  0.0220$ \\
			ST + Harmonics & $0.9008  \pm  0.0152$  &  $0.8645  \pm  0.0171$  &  $0.6665  \pm  0.0209$  &  $0.8634  \pm  0.0111$  &  $0.8768  \pm  0.0364$  &  $0.9043  \pm  0.0503$ \\
			GWST + Harmonics & $0.8726  \pm  0.0143$  &  $0.8461  \pm  0.0136$  &  $0.6218  \pm  0.0244$  &  $0.8495  \pm  0.0158$  &  $0.8832  \pm  0.0188$  &  $0.8594  \pm  0.0542$ \\ \hline
			
			\multicolumn{7}{|c|}{Group Lasso} \\ \hline
			GT + Harmonics & $0.9283  \pm  0.0130$ & $0.8949 \pm   0.0157$  &  $0.7416 \pm   0.0201$  &  $0.9318 \pm   0.0107$  &  $0.9008 \pm   0.0158$ &   $0.9261 \pm   0.0154$ \\
			MHWT + Harmonics &  $0.8523 \pm   0.0184$ &   $0.8001 \pm   0.0198$ &   $0.6811 \pm   0.0185$ &   $0.8195 \pm   0.0142$ &   $0.8015 \pm   0.0355$  &  $0.8304 \pm   0.0338$ \\
			ST + Harmonics &  $0.9629 \pm   0.0078$ &   $0.9402 \pm   0.0087$  &  $0.7069 \pm   0.0331$  &  $0.9566 \pm   0.0107$  &  $0.9422 \pm   0.0178$  &  $0.9358 \pm   0.0189$ \\
			GWST + Harmonics & $\boldsymbol{0.9752 \pm   0.0084}$ &   $\boldsymbol{0.9560 \pm   0.0096}$  &  $\boldsymbol{0.7589 \pm   0.0152}$  &  $\boldsymbol{0.9593 \pm   0.0098}$  &  $\boldsymbol{0.9588 \pm   0.0199}$ &   $\boldsymbol{0.9598 \pm   0.0116}$ \\ \bottomrule
			
		\end{tabular}
		\footnotetext{\scriptsize GT: Gabor transform. WT: Wavelet transform. ST: Stockwell transform. K-NN: K-nearest neighbours. LDC: linear discriminant classifier. QDC: quadratic discriminant classifier. SVM: support vector machine. ANN: artificial neural networks. $\mu$ and $\sigma$ are the mean and the standard deviation of the PQ classification performance.}
	\end{minipage}
	}
\end{table*}

To evaluate the performance of SLM for PQ classification, we first perform the synthesis step per each disturbance represented through GT, MHWT, ST, and GWST (combining GT, MHWT, ST), obtaining the set of coefficients $\{\Bbeta_p\}_{p=1}^P$. We then apply Group Lasso over the different representations, and compute the feature set $\{\Bx_p\}_{p=1}^P$ described in \ref{subsec:feature}. We train the different classifiers, and evaluate their performance over the test set as well as we describe in section \ref{subsubsec:classifier}

Table \ref{tab:result} shows the PQ classification performance over the test set. When Group Lasso is not applied (first four rows), ST and GWST show better results than GT and WT, independently of the classifier employed. When Group Lasso is applied (last four rows), we notice that for any particular representation, and any classifier, the accuracy obtained by additionally applying Group Lasso is higher, when compared to the same representation, and the same classifier used without Group Lasso. For example, when the representation is ST and the classifier is ANN, applying Group Lasso increases the performance by almost $4\%$. The improvement is even higher when the representation is GT, and the classifier is almost QDC. In this case the improvement is close to $33\%$.

For all the classifiers, accuracy results between both ST and GWST, using Group Lasso, are similar. This is particularly true for QDC and SVM. Due to the similarity of these accuracies, we apply a Wilcoxon rank-sum test (WRS) to evaluate the statistical significance of both results per classifier, concluding that the differences in performance are statistically significant.

Tables \ref{tab:None_NN_ST} and \ref{tab:GLasso_NN_ST} show the confusion matrices for GWST using 1-NN classifier without and with Group Lasso, respectively. From these tables we notice that without Group Lasso, misclassification occur in almost all the types of PQ disturbances, mainly in oscillatory distortions due to they require a high resolution in PQ representation both in time and frequency. On the other hand, the misclassification using Group Lasso are produced by the fact that similar frequency components appear in the different types of PQ disturbances (e.g. swell and sag, notch and oscillatory).

\begin{table}[h!]
	\begin{minipage}{\columnwidth}
		\centering%
		\caption{Confusion matrix for GWST and a 1-NN classifier. Group
			Lasso is not applied in this case.}
		{\scriptsize		
		\begin{minipage}{\textwidth}
			\centering
			\begin{tabular}{cccccccc}
				\toprule
				& $C_1$ & $C_2$ & $C_3$ & $C_4$ & $C_5$ & $C_6$ & $C_7$ \\ \midrule
				$C_1$ 	& \textbf{46} & 5 & 2 & 0 & 0 & 0 & 4 \\ 
				$C_2$ 		& 8 & \textbf{46} & 0 & 0 & 0 & 0 & 3 \\ 
				$C_3$ 		& 2 & 0 & \textbf{52} & 0 & 1 & 0 & 2 \\ 
				$C_4$ 	& 0 & 0 & 0 & \textbf{57} & 0 & 0 & 0 \\ 
				$C_5$ 		& 0 & 0 & 0 & 0 & \textbf{57} & 0 & 0 \\ 
				$C_6$ 	& 0 & 0 & 0 & 0 & 0 & \textbf{56} & 1 \\ 
				$C_7$ & 2 & 1 & 2 & 2 & 4 & 7 & \textbf{39} \\ \bottomrule		
			\end{tabular}
		\end{minipage}
		}
		\label{tab:None_NN_ST}
		\vspace{7pt}	
		
		\caption{Confusion matrix for GWST and a 1-NN classifier. Group
			Lasso is applied in this case.}
		{\scriptsize		
		\begin{minipage}{\textwidth}
			\centering	
			\begin{tabular}{cccccccc}
				\toprule
				& $C_1$ & $C_2$ & $C_3$ & $C_4$ & $C_5$ & $C_6$ & $C_7$ \\ \midrule
				$C_1$ 	& \textbf{57} & 0 & 0 & 0 & 0 & 0 & 0 \\ 
				$C_2$ 		& 0 & \textbf{56} & 1 & 0 & 0 & 0 & 0 \\ 
				$C_3$ 		& 0 & 2 & \textbf{55} & 0 & 0 & 0 & 0 \\ 
				$C_4$ 	& 0 & 0 & 0 & \textbf{57} & 0 & 0 & 0 \\ 
				$C_5$ 		& 0 & 0 & 0 & 0 & \textbf{57} & 0 & 0 \\ 
				$C_6$ 	& 0 & 0 & 0 & 0 & 0 & \textbf{57} & 0 \\ 
				$C_7$ & 0 & 0 & 0 & 0 & 2 & 0 & \textbf{55} \\ \bottomrule
			\end{tabular}
			\footnotetext{\centering \scriptsize $C_1$: harmonic. $C_2$: swell. $C_3$: sag. $C_4$: flicker. $C_5$: notch. $C_6$: impulsive transient. $C_7$: oscillatory transient.}
		\end{minipage}	
		}
		\label{tab:GLasso_NN_ST}
	\end{minipage}
\end{table}

\begin{figure}
	\centering
	\subfigure{\def\svgwidth{175pt} 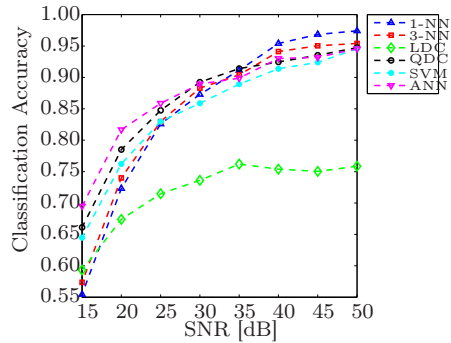}
	\caption{PQ disturbance classification performance using GWST plus Group Lasso under different levels of white Gaussian noise. NN: nearest neighbour. LDC: linear discriminant classifier. QDC: quadratic discriminant classifier. SVM: support vector machine. ANN: artificial
		neural network. }
	\label{eq:whitenoiseperformance}
\end{figure}

As an additional result, we evaluate the robustness of the framework proposed here in the presence of noise. In particular, we add white Gaussian noise to each PQ signal up until we get an specific Signal-to-Noise ratio (SNR). We then apply GWST plus Group Lasso over each signal, and run the classifiers twenty times, with different training and testing sets each time. Figure \ref{eq:whitenoiseperformance} shows the mean accuracy of the twenty repetitions obtained by each classifier, and for different SNR values. We notice that the classifiers show a reliable classification accuracy for SNR values greater than 30dB. For values below 30dB, the accuracy decreases due to the amount of noise, confusing ever more the waveforms among the PQ disturbance classification.  Notice also that for low SNR values, more sophisticated classifiers such as ANN, QDC, and SVM behave better than simpler classifiers like LDC or those based on nearest neighbours.

\section{Conclusion}
\label{sec:conclusion}

In this paper, we introduced overcomplete representations and sparse linear models for power quality disturbance classification. As an example of overcomplete representations, we combine the Gabor transform, the Wavelet transform with Mexican hat function, and the Stockwell transform, which are well known in the literature for PQ analysis. As an example of a sparse linear model, we use the Group Lasso assuming that each dictionary is a group in the sparse model.

Group Lasso selects automatically the most representative coefficients required to fully represent PQ disturbances, outperforming overcomplete representations. As we showed experimentally, this approach improves the performance of PQ disturbance classification for both linear and non-linear classifiers compared to methods without Group Lasso.

\section*{Acknowledgment}
We are thankful to the Department of Electrical Engineering, Universidad Tencol\'ogica de Pereira, for the technical support provided to carry out this research.

\appendix





\bibliographystyle{IEEEtran}
\bibliography{proyecto2}

\end{document}